\documentclass[pr,onecolumn,showpacs,superscriptaddress,10pt]{revtex4-2}
\usepackage[colorlinks,citecolor=blue,linkcolor=red,dvipdfm
]{hyperref}
\usepackage{amsmath,amssymb,graphicx}
\usepackage{times}

\begin{document}

\title{Vector solitons in nonlinear lattices with competing SPM and XPM terms%
}
\author{Liangwei Zeng}
\affiliation{School of Arts and Sciences, Guangzhou Maritime University,
Guangzhou 510725, China}

\author{Boris A. Malomed}
\affiliation{Department of Physical Electronics, School of Electrical Engineering, Faculty of Engineering,
and Center for Light-Matter Interaction, Tel Aviv University, P.O.B. 39040, Tel Aviv, Israel}
\affiliation{Instituto de Alta Investigaci\'{o}n, Universidad de
Tarapac\'{a}, Casilla 7D, Arica, Chile}

\author{Dumitru Mihalache}
\affiliation{Horia Hulubei National Institute of Physics and Nuclear
Engineering, 077125 Magurele, Bucharest, Romania}

\author{Xing Zhu}
\email{\underline{xingzhu@gzmtu.edu.cn}}
\affiliation{School of Arts and Sciences, Guangzhou Maritime University,
Guangzhou 510725, China}

\begin{abstract}
We demonstrate that two-component photonic nonlinear lattices with opposite signs of the out-of-phase spatially modulated SPM and XPM coefficients can support two types of optical vector solitons (VSs), the fundamental and mixed ones. The fundamental VSs are composed of two Gaussian-like components, while the mixed ones combine Gaussian-like and dipole components. The structures and stability of both types of the VSs are investigated. The VSs are stable, except for narrow intervals at small values of the propagation constant.
\end{abstract}

\maketitle

\section{Introduction}

The generation of solitons is a basic topic in many areas of nonlinear
physics \cite%
{Ichi,Newell,Zakh,REV1,Akhmed,Kiv,Dauxois,Laksh,Kuz,REV2,REV3,Greeks,multidim,El}%
. In particular, studies of solitons have drawn much interest in nonlinear
optics \cite{Abdullaev,GAP1,KA,OPT1,Chen,REV4,Tang,OPT3,Cui} and
Bose-Einstein condensates (BECs) \cite%
{Hulet,Brazh,BEC0,BEC1,Luca,BEC2,BEC3,BEC4}. Generally, the formation and
evolution of solitons is governed by diverse forms of the nonlinear Schr\"{o}%
dinger equation (NLSE). Various types of solitons have been produced with
the help of spatially periodic linear \cite{Brazh} and nonlinear \cite{REV2}
potentials, alias linear and nonlinear lattices. These are gap solitons,
populating finite bandgaps in the spectrum induced by the linear lattice
\cite{GAP2,GAP3,GAP5,GAP5a,GAP5b}, and solitons created by nonlinear
lattices, including fundamental \cite{NL3,AVK,NL1,GAP4}, dipole \cite{NL4},
multipole \cite{SATU9,SATU10}, dark \cite{NLD1,NLD2} and vortex \cite{NL6}
states.

Vector solitons (VSs) are modes which contain two or more components \cite%
{Laksh,Greeks,CS2,CS3,CS3a}, modeled by systems of coupled NLSEs. In
particular, VSs may carry different propagation constants (chemical
potentials, in terms of BEC) in their two components. Many types of VSs have
been investigated \cite{VTS1,VTS2,VTS3,VTS4,VTS5}, such as fundamental ones
\cite{VTS6}, dipoles \cite{VTS7,VTS8}, ``soliton molecules"
(bound states) \cite{VTS9}, ring solitons \cite{VTS10}, and vortices \cite%
{VTS11}. In particular, VSs of the gap type in systems combining the
saturable nonlinearity and linear-lattice potentials have been reported \cite%
{VTS12,VTS13}. Although the vector solitons in periodic linear potentials 
have been widely investigated, the ones under the action of periodically 
patterned self-phase modulation (SPM) and cross-phase modulation (XPM), 
in the absence of a linear lattice, call for more attention, especially to VSs built of components 
with different forms. Such nonlinear lattices and states supported by them 
may be realized in photorefractive crystals.

In this work, we study two-component VSs in nonlinear lattices (without any
linear potential). Two types of VSs are produced, namely, the fundamental
and mixed ones. The former species is built as a single-peak mode with Gaussian-like shapes of its components, while the mixed solitons combine fundamental and dipole
components.

The presentation is arranged as follows. We report the model and
linear-stability analysis of the VSs in Section \ref{sec2}, where we also
provide a qualitative explanation of the possibility of the formation of
stable VSs in the system. Numerical results\ are reported in Section \ref%
{sec3}, which is divided in two parts, that address the fundamental and
mixed VSs, respectively. The paper is concluded by Section \ref{sec4}.

\section{The model}

\label{sec2}

The propagations of optical VSs in the nonlinear lattice are modeled by the
system of coupled NLSEs, written in terms of the scaled propagation distance
$z$ and transverse coordinate $x$:
\begin{eqnarray}
i\frac{{\partial E_{1}}}{{\partial z}} &=&-\frac{1}{2}\frac{\partial
^{2}E_{1}}{\partial x^{2}}+\left[ G(x)|E_{1}|^{2}+\Gamma (x)|E_{2}|^{2}%
\right] E_{1},  \notag \\
i\frac{{\partial E_{2}}}{{\partial z}} &=&-\frac{1}{2}\frac{\partial
^{2}E_{2}}{\partial x^{2}}+\left[ G(x)|E_{2}|^{2}+\Gamma (x)|E_{1}|^{2}%
\right] E_{2}.  \label{NLSE}
\end{eqnarray}%
Variables $E_{1,2}$ are the slowly varying amplitudes of the two components,
which may represent orthogonal polarizations of light or optical waves with
different carrier wavelengths. The nonlinear lattice is introduced by
adopting spatially-modulated expressions,%
\begin{equation}
G(x)=-\mathrm{cos}^{2}x,~\Gamma (x)=\gamma \mathrm{sin}^{2}x,  \label{GGamma}
\end{equation}%
for the local SPM and XPM coefficients, with $\gamma >0$, which makes the
SPM and XPM signs opposite, corresponding to the competition between the
focusing and defocusing nonlinearities. The lattice period in Eq. (\ref%
{GGamma}) is fixed to be $\pi $ by means of scaling. We consider the systems
with $\gamma =2/3$ and $2$, which correspond, respectively, to the XPM
interaction between mutually orthogonal linear polarizations of light, or
the interaction between left- and right-twisted circular polarizations of
light, as well as the between optical waves carried by different wavelengths
\cite{KA}. In fact, values of $\gamma $ between $2/3$ and $\ 2$ produce
results which are quite similar to those reported below. The opposite SPM
and XPM signs are possible in photonic crystals \cite{Panoiu,Bianc}, in
which the spatial modulation can be fabricated by means of available
technologies \cite{hetero,Notomi}. The system of equations \ref{NLSE})
conserves three dynamical invariants: powers of the two components,%
\begin{equation}
P_{1,2}=\int_{-\infty }^{+\infty }\left\vert E_{1,2}(x)\right\vert ^{2}dx,
\label{P}
\end{equation}%
and the Hamiltonian,%
\begin{equation}
H=\int_{-\infty }^{+\infty }\left\{ \frac{1}{2}\sum_{n=1,2}\left[ \left\vert
\frac{\partial E_{n}}{\partial x}\right\vert +G\left\vert E_{n}\right\vert
^{4}\right] +\Gamma |E_{1}|^{2}|E_{2}|^{2}\right\} dx.  \label{H}
\end{equation}

Stationary solutions for the VSs are produced by substituting
\begin{equation}
E_{1,2}=U_{1,2}(x)\mathrm{exp}(ib_{1,2}z)  \label{12}
\end{equation}%
in Eq. (\ref{NLSE}), where $b_{1,2}>0$ are propagation constants of the two
components. This leads to the system of stationary equations for real wave
functions $U_{1,2}$,
\begin{eqnarray}
-b_{1}U_{1} &=&-\frac{1}{2}\frac{d^{2}U_{1}}{dx^{2}}+\left[
G(x)U_{1}^{2}+\Gamma (x)U_{2}^{2}\right] U_{1},  \notag \\
-b_{2}U_{2} &=&-\frac{1}{2}\frac{d^{2}U_{2}}{dx^{2}}+\left[
G(x)U_{2}^{2}+\Gamma (x)U_{1}^{2}\right] U_{2}.  \label{NLSES}
\end{eqnarray}

In the spatially uniform system, with constant coefficients $G<0$ and $%
\Gamma >0$, VSs in the form of a bound state of two fundamental components $%
E_{1}$ and $E_{2}$ cannot be stable, as $\Gamma >0$ implies repulsion
between the components, hence the two-component soliton would\ be unstable
against spontaneous splitting into separating single-component ones.
Nevertheless, the nonlinear lattice structure, adopted in Eq. (\ref{GGamma}%
), makes the existence of a stable two-component mode possible. To
demonstrate this possibility in an approximate form, one can treat the XPM
terms in Eqs. (\ref{NLSE}) and (\ref{NLSES}) as small perturbations, which
is possible if $\gamma $ in Eq. (\ref{GGamma}) is a small coefficient, or
the power of one soliton is much smaller than the other's, $P_{2}\ll P_{1}$
(in other words, $b_{2}\ll b_{1}$ in Eqs. (\ref{NLSES})). Then, in the
zeroth-order approximation, the $E_{1}$ component is shaped as a fundamental
soliton, $\left( U_{1}\right) _{\mathrm{sol}}=2\sqrt{b_{1}}\mathrm{sech}\left(
\sqrt{2b_{1}}x\right)$, if the spatially-modulated SPM coefficient $G(x)$
is replaced by its average value, $\bar{G}=-1/2$ (see Eq. (\ref{GGamma})).
Finally, the XPM term in the second equation of system (\ref{NLSE}) gives
rise to an effective potential for field $E_{2}$:
\begin{equation}
W_{\mathrm{eff}}(x)=4b_{1}\gamma \frac{\sin ^{2}x}{\cosh ^{2}\left(
\sqrt{2b_{1}}x\right) },  \label{W}
\end{equation}
where $\Gamma (x)$ is substituted as per Eq. (\ref{GGamma}). Expression (\ref{W})
describes a \textit{potential quasi-well} centered at $x=0$. In the linear limit,
the quasi-well supports leaky (gradually decaying) bound states, as the potential,
vanishing at $|x|\to \infty$, is positive at all $x$. Nevertheless, the inclusion of the
self-focusing nonlinearity readily transforms the leaky bound states into truly bound ones
\cite{Moiseyev}. In other words, the potential quasi-well may provide stable
trapping of a soliton in the $E_{2}$ component, thus building a genuine robust
VS.

As concerns the bound state of the fundamental mode in $E_{1}$ and the
dipole one in $E_{2}$, the latter component may be considered, in a crude
approximation, as a pair of out-of-phase solitons trapped in the nonlinear
lattice induced by coefficient $G(x)$ (see Eq. (\ref{GGamma})), which
interact with the fundamental mode in component $E_{1}$ through the
effective potential (\ref{W}).

\begin{figure}[tbp]
	\begin{center}
		\includegraphics[width=0.8\textwidth]{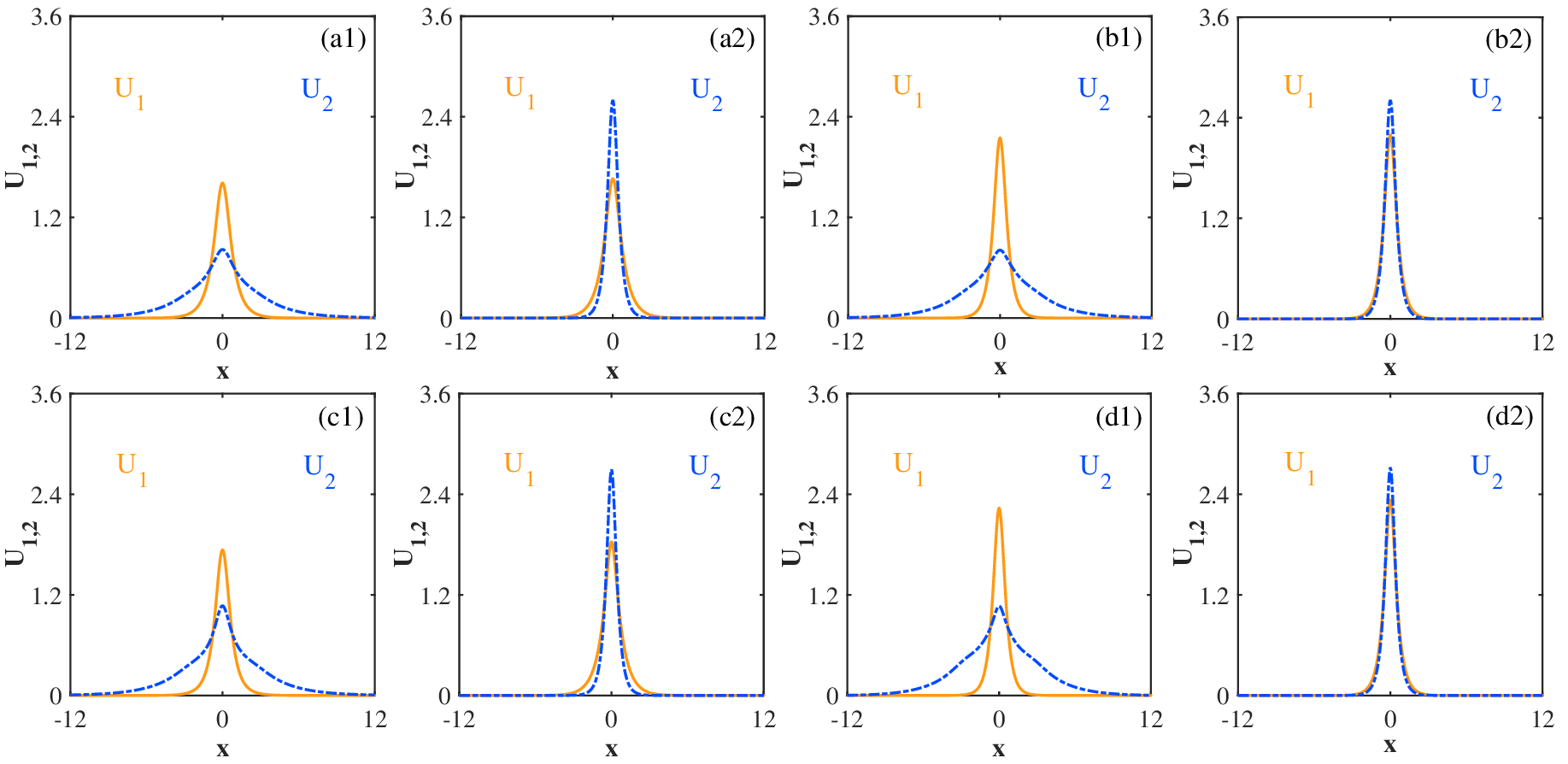}
	\end{center}
	\caption{Profiles of fundamental VSs in the first and second
		components (yellow and blue lines, respectively) for $\protect\gamma =2/3$:
		(a1) unstable one at $b_{1}=1$, $b_{2}=0.1$; (a2) stable one at $b_{1}=1$, $b_{2}=3$; (b1) unstable one at $b_{1}=2$, $b_{2}=0.1$; (b2) stable one at $b_{1}=2$, $b_{2}=3$. Panels (c1)--(d2): the same as in
		(a1)--(b2) but for $\protect\gamma =2$.}
	\label{fig1}
\end{figure}

\begin{figure}[tbp]
\begin{center}
\includegraphics[width=0.8\textwidth]{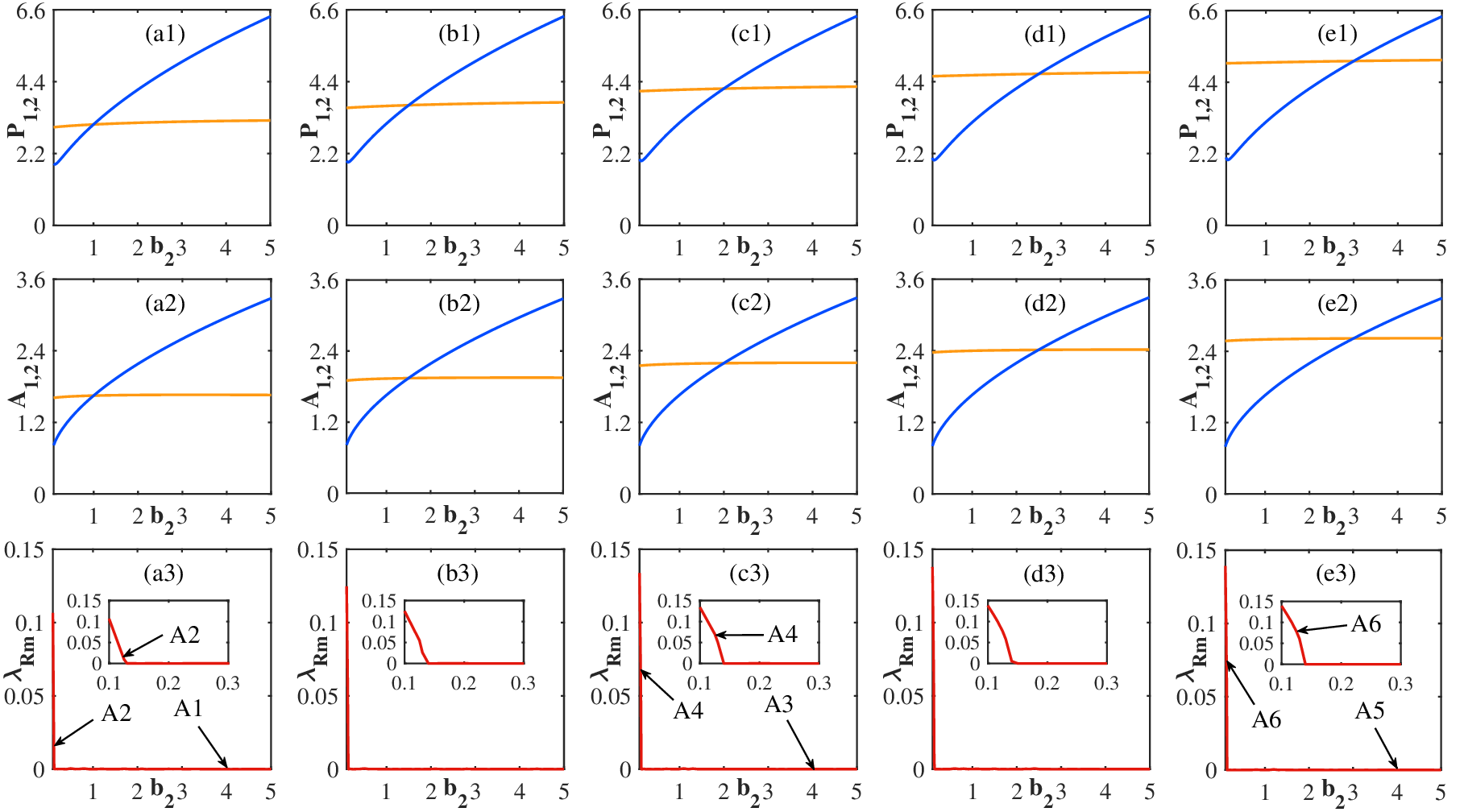}
\end{center}
\caption{Powers $P_{1,2}$ (top row), amplitudes $A_{1,2}$ (middle row) ,and
maximum real parts of eigenvalues $\protect\lambda _{\mathrm{Rm}}$ (bottom
row) of fundamental VSs versus the propagation constant of the second
component, $b_{2}$, for $\protect\gamma =2/3$: (a1)--(a3) with $b_{1}=1$;
(b1)--(b3) with $b_{1}=1.5$; (c1)--(c3) with $b_{1}=2$; (d1)--(d3) with $%
b_{1}=2.5$; (e1)--(e3) with $b_{1}=3$. The yellow and blue lines pertain to
the first and second components, respectively.}
\label{fig2}
\end{figure}

To explore the stability of the VSs against small perturbations, we disturb
the VS solution as
\begin{equation}
E_{1,2}(x,z)=[U_{1,2}(x)+p_{1,2}(x){e}^{\lambda z}+q_{1,2}^{\ast }(x){e}%
^{\lambda ^{\ast }z}]{e}^{ib_{1,2}z},  \label{PERT}
\end{equation}%
where $\lambda $ (which may be complex) is the instability growth rate
corresponding to the perturbation eigenmode with components $p_{1,2}(x)$ and
$q_{1,2}^{\ast }(x)$, the asterisk standing for the complex conjugate. The
substitution of ansatz (\ref{PERT}) into Eq. (\ref{NLSE}) and linearization
with respect to the perturbation leads to the following eigenvalue problem
for $\lambda $:
\begin{equation}
i\lambda \left(
\begin{array}{c}
p_{1} \\
q_{1} \\
p_{2} \\
q_{2}%
\end{array}%
\right) =\left(
\begin{array}{cccc}
\hat{L}_{1} & \hat{W}_{1} & \hat{W}_{2} & \hat{W}_{2} \\
-\hat{W}_{1} & -\hat{L}_{1} & -\hat{W}_{2} & -\hat{W}_{2} \\
\hat{W}_{2} & \hat{W}_{2} & \hat{L}_{2} & \hat{W}_{3} \\
-\hat{W}_{2} & -\hat{W}_{2} & -\hat{W}_{3} & -\hat{L}_{2}%
\end{array}%
\right) \left(
\begin{array}{c}
p_{1} \\
q_{1} \\
p_{2} \\
q_{2}%
\end{array}%
\right) ,  \label{LSA}
\end{equation}%
where
\begin{equation}
\begin{aligned} \hat{L}_{1}=&-(1/2)(\partial^2/\partial
x^2)+b_1+2GU_1^2+\Gamma U_2^2, \\ \hat{L}_{2}=&-(1/2)(\partial^2/\partial
x^2)+b_2+2GU_2^2+\Gamma U_1^2, \\ \hat{W}_{1}=&GU_1^2, \hat{W}_{2}=\Gamma
U_1U_2, \hat{W}_{3}=GU_2^2. \end{aligned}
\end{equation}%
Obviously, the stationary solutions are stable if Re$(\lambda )=0$ for all
eigenvalues.

Below, the stationary VS solutions of Eq. (\ref{NLSES}) are obtained by
means of the modified squared-operator method \cite{MSOM}, The stability of
VSs is identified by the spectrum of eigenvalues $\lambda $ and is
corroborated by numerical simulations of the perturbed evolution, which are
performed using the standard finite-difference method. In this work, the computational domain for the fundamental VSs is $x\in[-15, 15]$, while the one for the mixed VSs is $x\in[-20, 20]$. The spatial and propagation steps are $dx=0.05$ and $dz=0.00025$, respectively. The convergence tolerance is $10^{-10}$. We use the Laplacian-zero (neglecting the term of Laplacian second derivative) \cite{L0B} as the boundary condition in this work. The perturbation amplitude is $1\%$ of the stationary solution. And we employ the Fourier collocation method \cite{FCM} as the eigensolver.

\section{Numerical results}

\label{sec3}

In this section, we focus on the numerical results for the VSs, including
profiles, power, amplitude, stability and perturbed propagations of the
solitons. Two types of VSs are investigated in this section, namely, the
fundamental and mixed VSs. In the former case, both components are
fundamental modes. In the mixed VS, one component remains fundamental, while
the other one features a dipole shape. The results for the fundamental and
mixed VSs are presented here in subsections \ref{subsec31} and \ref{subsec32}%
, respectively.

\begin{figure}[tbph]
\begin{center}
\includegraphics[width=0.8\textwidth]{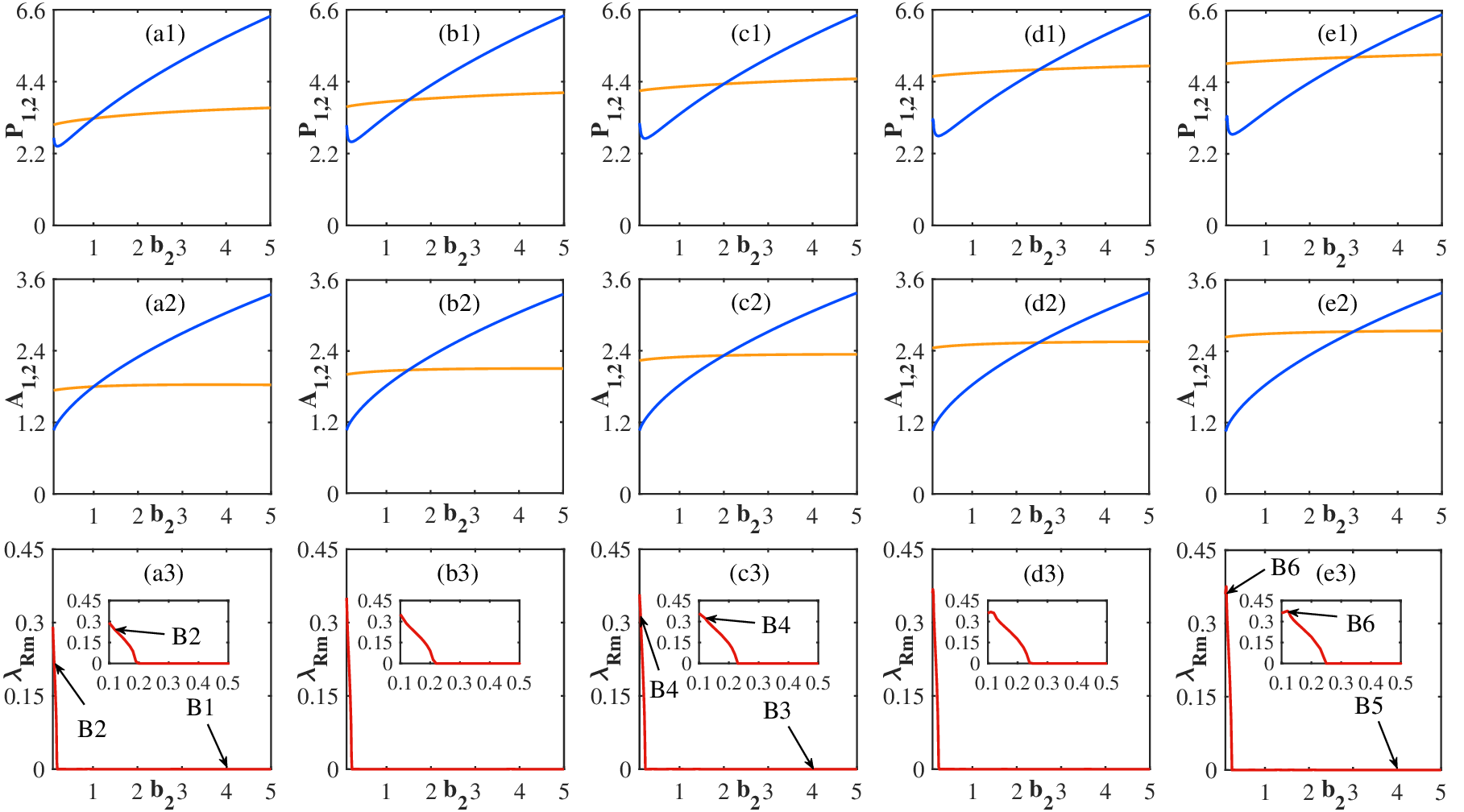}
\end{center}
\caption{The same as in Fig. \protect\ref{fig2}, but with $\protect\gamma =2$.}
\label{fig3}
\end{figure}

\subsection{Fundamental VSs}

\label{subsec31}

For the physically relevant value of the XPM coefficient $\gamma =2/3$,
typical profiles of unstable fundamental VSs with widely different propagation
constants of the first and second components, \textit{viz}., $%
b_{1}=1,b_{2}=0.1$ and $b_{1}=2,b_{2}=0.1$ are presented in Figs. \ref{fig1}%
(a1) and (b1), respectively. In this case, the second equation in system (%
\ref{NLSES}) suggests that the width of the second component scales as $%
b_{2}^{-1/2}$, i.e., it is relatively large when $b_{2}$ is small and,
naturally, the amplitude of this component is smaller than its counterpart
in the first component, $A_{2}<A_{1}$. On the other hand, when the values of
the propagation constants take values on the same order of magnitude,
\textit{viz}., $b_{1}=1,b_{2}=3$ and $b_{1}=2,b_{2}=3$ in Figs. \ref{fig1}%
(a2) and (b2), respectively, the widths of the two components are nearly
equal, and the amplitude $A$ is larger for larger $b$. Figures \ref{fig1}%
(c1)--(d2) present similar results, but for the other physically relevant
value of the XPM coefficient, $\gamma =2$ (recall that it corresponds to the
circular polarizations of light or different wavelengths).

\begin{figure}[tbp]
\begin{center}
\includegraphics[width=0.8\textwidth]{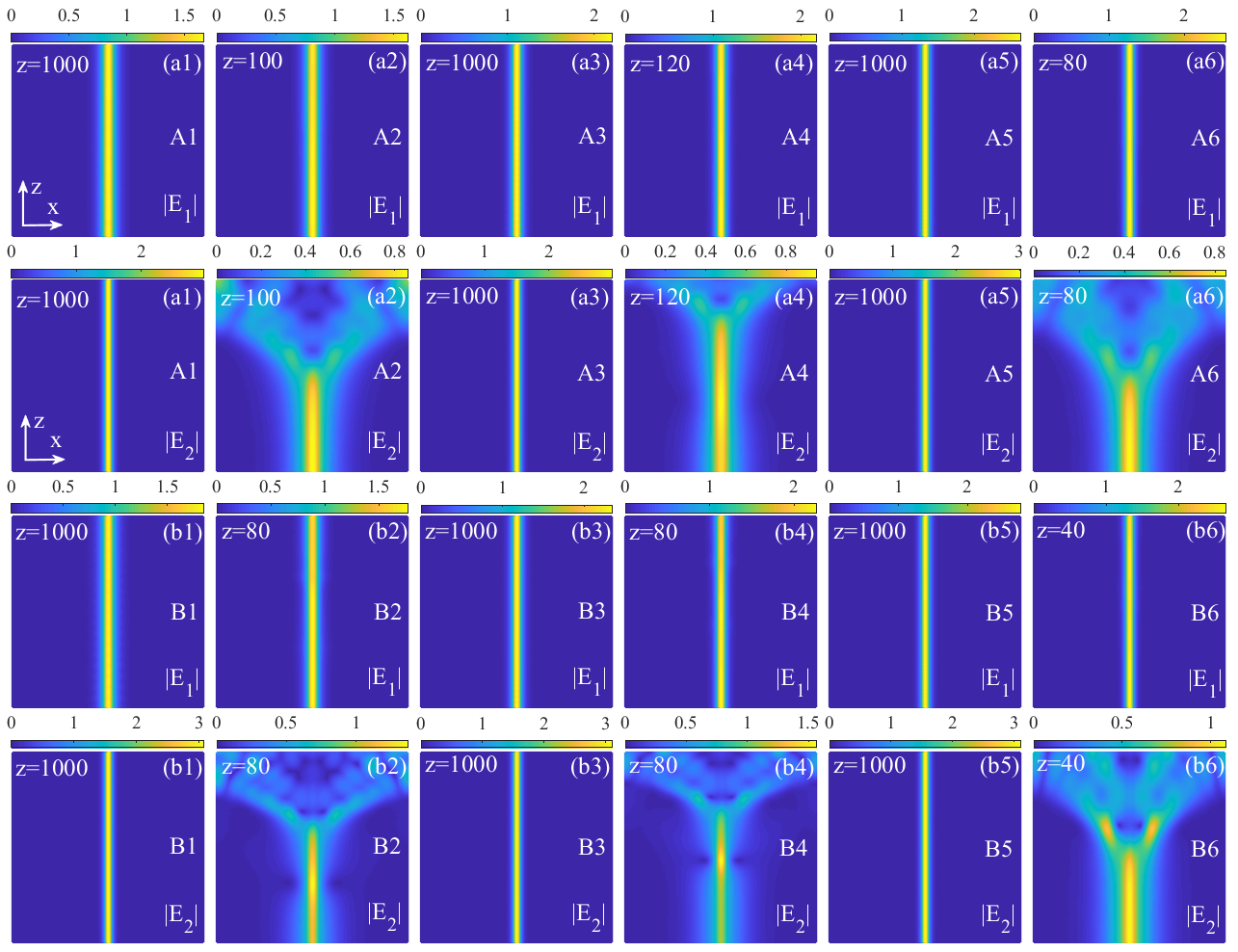}
\end{center}
\caption{Perturbed propagations of fundamental VSs at $\protect\gamma =2/3$:
(a1) with $b_{1}=1$, $b_{2}=4$; (a2) with $b_{1}=1$, $b_{2}=0.125$; (a3)
with $b_{1}=2$, $b_{2}=4$; (a4) with $b_{1}=2$, $b_{2}=0.125$; (a5) with $%
b_{1}=3$, $b_{2}=4$; (a6) with $b_{1}=3$, $b_{2}=0.125$. Panels (b1) - (b6)
show the same for $\protect\gamma =2$. In all panels, the simulation domain
is $|x|\leq 15$.}
\label{fig4}
\end{figure}

Next, we systematically address the power, amplitude and stability of the
fundamental-VS family. These characteristics for the XPM coefficient $\gamma
$ are presented, severally, in the top, middle and bottom rows of Figs. %
\ref{fig2} and \ref{fig3}, for $\gamma =2/3$ and $2$, respectively, as
functions of $b_{2}$, at several fixed values of the propagation constant of
the first component, \textit{viz}., $b_{1}=1,1.5,2,2.5$, It is seen that
power $P_{2}$ always increases with the increase of $b_{2}$, while power $P_{1}$ increases very slowly, almost keeping a constant value.
Further, the amplitude of the second component, $A_{2}$, increases with the
increase of $b_{2}$, while $A_{1}$ remains almost constant.

The maximum real part $\lambda _{\mathrm{Rm}}$ of the instability growth
rate [given by Eqs. (\ref{LSA})] versus $b_{2}$ with the values of $b_{1}=1,1.5,2,2.5,3$
of fundamental VSs at $\gamma =2/3$ are portrayed in Figs. \ref{fig2}%
(a3)--(e3), respectively. According to Figs. \ref{fig2}(a3)--(e3), these
fundamental VSs become unstable ($\lambda _{\mathrm{Rm}}>0$) when $b_{2}$ is
very small (in the left parts of the panels). And the curves of $\lambda _{\mathrm{Rm}}$ versus $b_2$ for such unstable VSs can be clearly seen in the subgraphs of these panels. The perturbed propagations of the fundamental VSs labelled by
A1--A6 are reported in Figs. \ref{fig4}(a1)--(a6), respectively.

The power, amplitude and linear stability analysis of the fundamental VSs at
$\gamma =2$ are presented in the top, middle and bottom rows of Fig. \ref%
{fig3}, respectively. Specifically, the power of both components $P_{1}$ and
$P_{2}$ versus $b_{2}$ with values of $b_{1}=1,1.5,2,2.5,3$ are reported
in Figs. \ref{fig3}(a1)--(e1) from which one can see that $P_{1}$ (yellow line)
and $P_{2}$ (blue line) always increase with the increase of $b_{2}$.
And the increasing speed of $P_{1}$ versus $b_{2}$ at $\gamma =2$ is slower
than the one of $P_{2}$. In addition, $P_{1}$ increase with the increase of $%
b_{1}$ obviously. Next we present the amplitude of both components versus $%
b_{2}$ in Figs. \ref{fig3}(a2)--(e2) where both $A_{1}$ and $A_{2}$ increase
with the increase of $b_{2}$ and the increasing speed of $A_{2}$ versus $%
b_{2}$ is faster than the one of $A_{1}$. Further, we can also find that $%
A_{1}$ increase with $b_{1}$ obviously.

\begin{figure}[tbp]
\begin{center}
\includegraphics[width=0.8\textwidth]{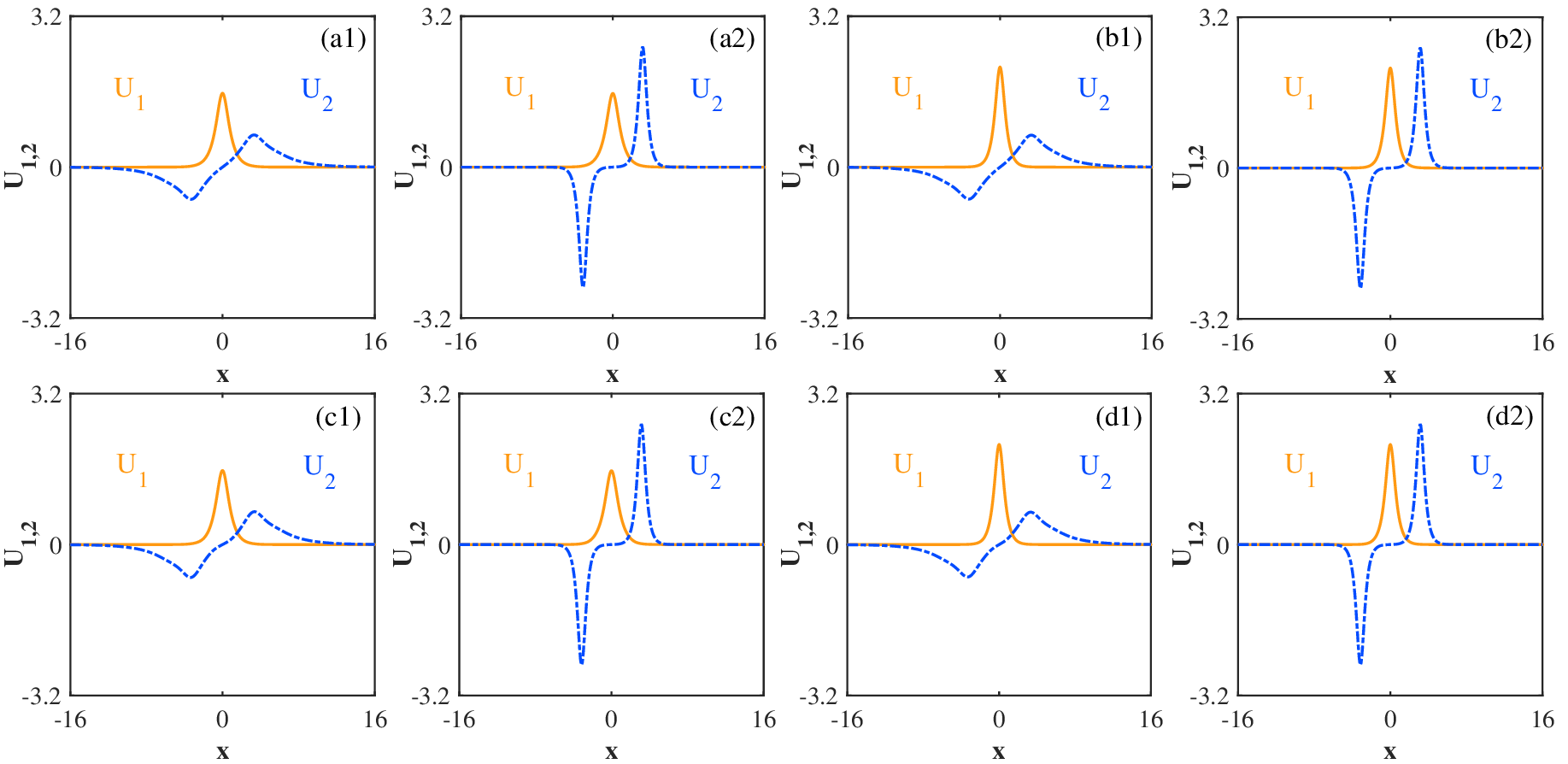}
\end{center}
\caption{Profiles of mixed VSs in the first and second components
(yellow and blue lines, respectively) with the XPM parameter $\protect\gamma %
=2/3$: (a1) unstable one at $b_{1}=1$, $b_{2}=0.1$; (a2) stable one at $b_{1}=1$, $b_{2}=3$; (b1) unstable one at $b_{1}=2$, $b_{2}=0.1$; (b2) stable one at $b_{1}=2$, $b_{2}=3$. Panels (c1)--(d2): the same as
(a1)--(b2), but for $\protect\gamma =2$.}
\label{fig5}
\end{figure}

\begin{figure}[tbp]
\begin{center}
\includegraphics[width=0.8\textwidth]{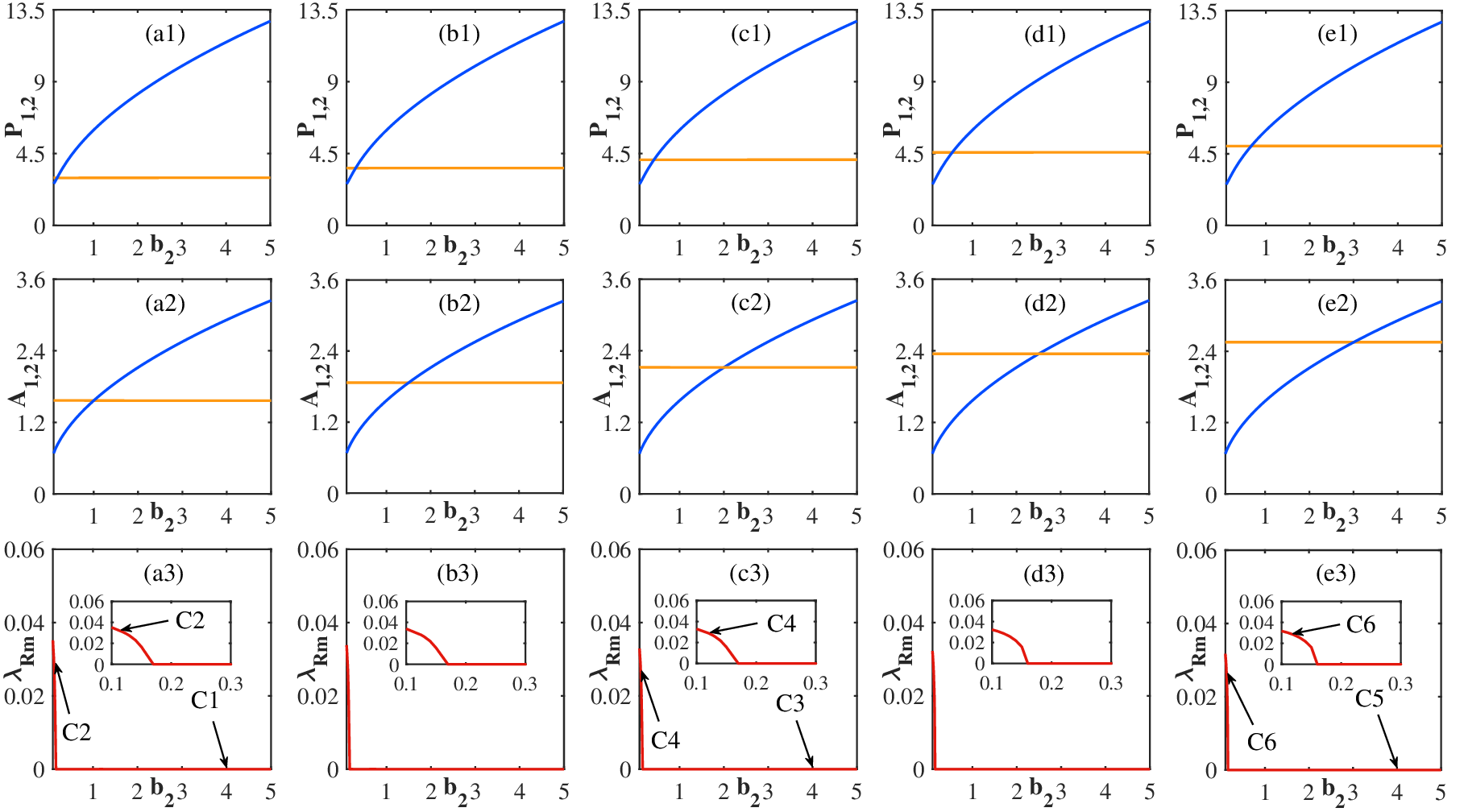}
\end{center}
\caption{Powers $P_{1,2}$ (the top row), amplitudes $A_{1,2}$ (the middle
row), and the maximum real part of eigenvalues $\protect\lambda _{\mathrm{Rm}%
}$ (the bottom row) of mixed VSs, for $\protect\gamma =2/3$, versus the
propagation constant $b_{2}$ of the dipole component $b_{2}$: (a1)--(a3) for
$b_{1}=1$; (b1)--(b3) for $b_{1}=1.5$; (c1)--(c3) for $b_{1}=2$; (d1)--(d3)
for $b_{1}=2.5$; (e1)--(e3) for $b_{1}=3$. Yellow and blue lines display the
results of the fundamental and dipole components, respectively.}
\label{fig6}
\end{figure}

The dependence of the maximum real part $\lambda _{\mathrm{Rm}}$ of the
instability growth rate [produced by the numerical solution of Eqs. (\ref%
{LSA})] versus $b_{2}$ with the values on $b_{2}$ demonstrates that the
fundamental VSs are unstable ($\lambda _{\mathrm{Rm}}>0$) only at very small
values of $b_{2}$. To verify the linear-stability predictions, simulations
of the perturbed propagation of the fundamental VSs are presented in Fig. %
\ref{fig4}. As expected, the VSs with $\lambda _{\mathrm{Rm}}=0$, labelled
A1, A3, A5, B1, B3, B5 in Fig. \ref{fig4} remain stable up to the largest
propagation distance tested here, $z=1000$. On the other hand, the
fundamental VSs labelled A2, A4, A6, B2, B4, B6, which correspond to $%
\lambda _{\mathrm{Rm}}>0$, are indeed unstable, exhibiting sudden decay of
the amplitude of the second component, thus spontaneously transforming into obviously stable
single-component solitons.

\subsection{Mixed VSs}

\begin{figure}[tbp]
\begin{center}
\includegraphics[width=0.8\textwidth]{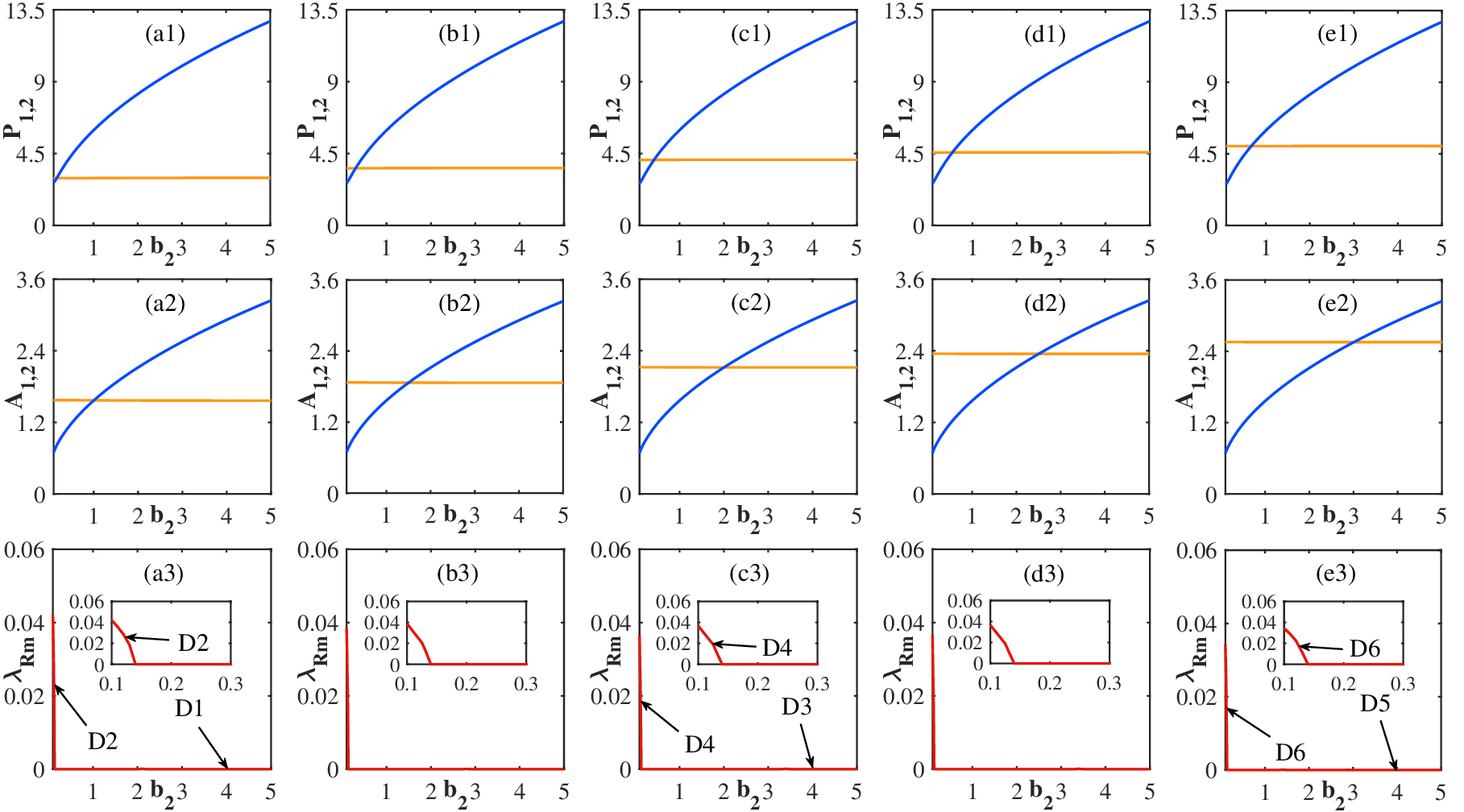}
\end{center}
\caption{The same as in Fig. \protect\ref{fig6}, but for $\protect\gamma =2$.
}
\label{fig7}
\end{figure}

\begin{figure}[tbp]
\begin{center}
\includegraphics[width=0.8\textwidth]{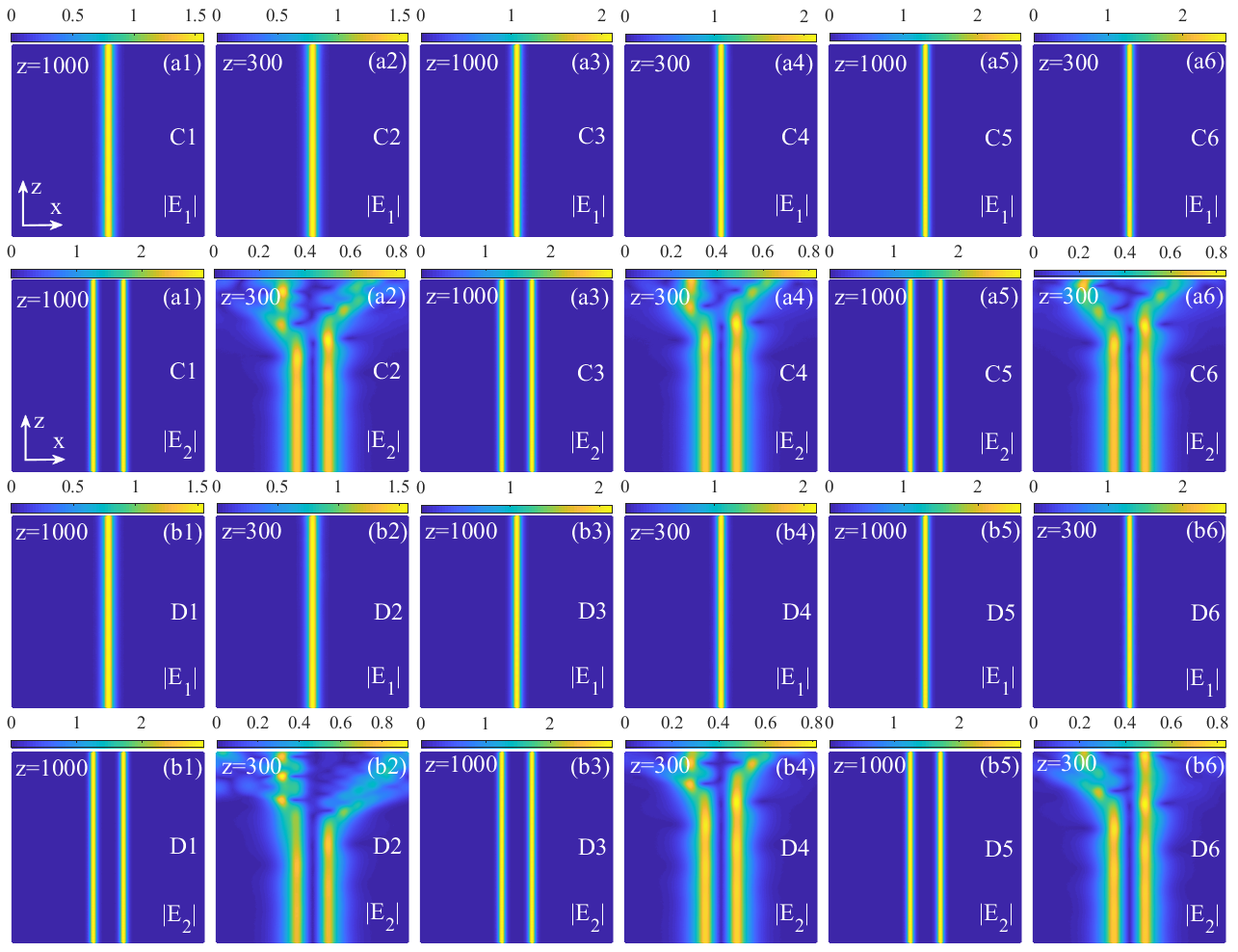}
\end{center}
\caption{The perturbed propagations of mixed VSs for $\protect\gamma =2/3$:
(a1) with $b_{1}=1$, $b_{2}=4$; (a2) with $b_{1}=1$, $b_{2}=0.125$; (a3)
with $b_{1}=2$, $b_{2}=4$; (a4) with $b_{1}=2$, $b_{2}=0.125$; (a5) with $%
b_{1}=3$, $b_{2}=4$; (a6) with $b_{1}=3$, $b_{2}=0.125$. Panels (b1) - (b6)
show the same for $\protect\gamma =2$. \ In all panels, the simulation
domain is $|x|\leq 20.$}
\label{fig8}
\end{figure}

\begin{figure}[tbp]
	\begin{center}
		\includegraphics[width=0.8\textwidth]{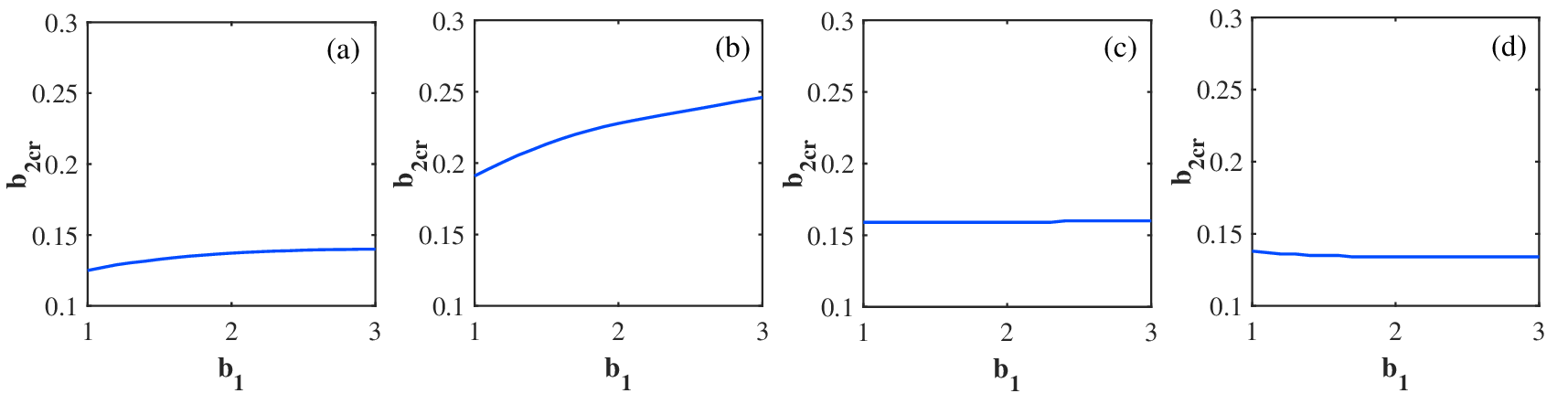}
	\end{center}
	\caption{The critical value of $b_2$ versus $b_1$ for these soliton families: (a) fundamental VSs at $\gamma=2/3$; (b) fundamental VSs at $\gamma=2$; (c) mixed VSs at $\gamma=2/3$; (d) mixed VSs at $\gamma=2$.}
	\label{fig9}
\end{figure}

\label{subsec32}

Recall that the mixed VSs are composed of two different components,
fundamental and dipole ones. Profiles of the mixed VSs with small and
relatively large values of the dipole's propagation constants $b_{2}$ are
presented in Figs. \ref{fig5}(a1,b1,c1,d1) and (a2,b2,c2,d2), respectively.
In the former case, the dipole's amplitude is small, and its shape is
smooth, while in the latter case the amplitude is much higher, and the
dipole effectively splits in a pair of separated peaks with opposite signs.

Dependences of the power, amplitude and linear stability of the mixed-VS
family on the propagation constant $b_{2}$ of the dipole component, for $%
\gamma =2/3$ and $\gamma =2$, are plotted in the top, middle, and bottom
rows of Figs. \ref{fig6} and \ref{fig7}, respectively. In both figures, the
dependences are produced for fixed values of the propagation constant of the
fundamental component, $b_{1}=1.0,1.5,2.0,2.5,3.0$, in columns (a)--(e),
respectively.

In the same Figs. \ref{fig6} and \ref{fig7}, panels (a3)-(e3) display
dependences of the maximum real part $\lambda _{\mathrm{Rm}}$ of the
instability growth rate versus $b_{2}$. It is seen that, as well as the
fundamental VSs [see panels (a3)-(e3) in Figs. \ref{fig2} and \ref{fig3}],
the mixed ones have instability intervals only at very small values of $b_{2}
$ (in the left parts of the panels). We also present the curves of $\lambda _{\mathrm{Rm}}$ versus $b_2$ with the small value of $b_2$ (unstable mixed VSs) in the subgraphs of the panels (a3)--(e3).

The predictions for the stability of mixed VS, produced by the
linear-stability analysis, is verified by direct simulations of the
perturbed propagation in Fig. \ref{fig8}. It is seen that the mixed VSs
labelled by C1, C3, C5, D1, D3, D5 in Fig. \ref{fig8} are stable, while ones
labelled by C2, C4, C6, D2, D4, D6 are unstable. Similar to their
counterparts of the fundamental type (see Fig. \ref{fig4}), the dipolar
components of the unstable mixed VSs get spontaneously destroyed,  while the
fundamental components survive as single-component solitons.

Finally, to investigate the stability domains of the soliton families, relations between the critical value of $b_2$ ($b_{2cr}$) and $b_1$, for both fundamental and mixed VSs, are presented in Fig. \ref{fig9}. Specifically, the curves of $b_{2cr}(b_1)$ for the fundamental VSs, with $\gamma=2/3$ and $\gamma=2$, are plotted in Figs. \ref{fig9}(a) and \ref{fig9}(b), while the ones for mixed VSs, with $\gamma=2/3$ and $\gamma=2$, are plotted in Figs. \ref{fig9}(c) and \ref{fig9}(d). Obviously, in panel (b) the value of $b_{2cr}$ is sensitive to the value of $b_1$, while the variation of $b_{2cr}$ with the increase of $b_1$ is quite weak in other panels.

\section{Conclusion}

\label{sec4}

In this work, we have demonstrated that nonlinear photonic lattices (in the
absence of the linear potential) with opposite signs of the SPM and XPM
terms sustain two types of optical VSs (vector solitons), \textit{viz}., the
fundamental and mixed ones. The fundamental VSs are built of two fundamental
components, while the mixed ones combined fundamental and dipole components.
The shapes, powers, amplitudes and linear stability of both VS types are
systematically investigated. In particular, the dependences of these
characteristics of the mixed VS family on the propagation constant of the
dipolar component are reported. The stability domains of these VSs,
predicted by the linear-stability analysis, have been verified by direct
simulations of the perturbed propagation. The VSs are unstable only at very
small values of the propagation constant of one component (the dipolar one,
in the case of mixed VSs).

\section*{Funding}
National Natural Science Foundation of China (Grants No. 62205224 and No. 11774068), Guangdong Basic and Applied Basic Research Foundation (Grants No. 2026A1515010001 and No. 2023A1515010865), Guangdong Association for Science and Technology Young Talent Development Program (Grant No. SKXRC2026684), Characteristic Innovation Projects of Guangdong Provincial General Colleges and Universities (2025KTSCX104), and Guangzhou Science and Technology Plan Project (Grant No. 2025A04J4068).

\section*{Data Availability Statement}
 Data will be made available on a reasonable request.

\section*{Conflict of interest}
 The authors declare that they have no conflict of interest.


\begin{thebibliography}{99}
\bibitem{Ichi} Y.H. Ichikawa, Topics on Solitons in Plasmas. Phys. Scripta \textbf{20}, 296--305 (1979)

\bibitem{Newell} A. Newell, \textit{Solitons in Mathematics and Physics} (SIAM, Philadelphia, 1985)

\bibitem{Zakh} E.A. Kuznetsov, A.M. Rubenchik, V.E. Zakharov, Soliton stability in plasmas and hydrodynamics. Phys. Rep. \textbf{142}, 103--165 (1986)

\bibitem{REV1} Y.S. Kivshar and B.A. Malomed, Dynamics of solitons in nearly integrable systems. Rev. Mod. Phys. \textbf{61}, 763--915 (1989)

\bibitem{Akhmed} N. Akhmediev, A. Ankiewicz, \textit{Solitons: Nonlinear Pulses and Beams} (Chapman \& Hall, London, 1997)

\bibitem{Kiv} Y.S. Kivshar, D. E. Pelinovsky, Self-focusing and transverse instabilities of solitary waves. Phys. Rep. \textbf{331}, 117--195 (2000)

\bibitem{Dauxois} T. Dauxois, M. Peyrard, \textit{Physics of Solitons} (Cambridge University Press, Cambridge, 2006)

\bibitem{Laksh} M. Vijayajayanthi, T. Kanna, M. Lakshmanan, Multisoliton solutions and energy sharing collisions in coupled nonlinear Schr\"{o}dinger equations with focusing, defocusing and mixed type nonlinearities. Eur. Phys. J. Spec. Top. \textbf{173}, 57--80 (2009)

\bibitem{Kuz} E.A. Kuznetsov, F. Dias, Bifurcations of solitons and their stability. Phys. Rep. \textbf{507}, 43--105 (2011)

\bibitem{REV2} Y.V. Kartashov, B.A. Malomed, L. Torner, Solitons in nonlinear lattices. Rev. Mod. Phys. \textbf{83}, 247--305 (2011)

\bibitem{REV3} V.V. Konotop, J. Yang, D.A. Zezyulin, Nonlinear waves in $\mathcal{PT}$-symmetric systems. Rev. Mod. Phys. \textbf{88}, 035002 (2016)

\bibitem{Greeks} P. G. Kevrekidis, D. J. Frantzeskakis, Solitons in coupled nonlinear Schr\"{o}dinger models: A survey of recent developments. Rev. Phys. \textbf{1}, 140--153 (2016)

\bibitem{multidim} B.A. Malomed, \textit{Multidimensional Solitons} (AIP Publishing, Melville, NY, 2022)

\bibitem{El} P. Suret, S. Randoux, A. Gelash, D. Agafontsev, B. Doyon, G. El, Soliton gas: Theory, numerics, and experiments. Phys. Rev. E \textbf{109}, 061001 (2024) (perspective)

\bibitem{Abdullaev} F. Abdullaev, S. Darmanyn, P. Khabibullaev, \textit{Optical Solitons} (Springer-Verlag, Berlin, 1993)

\bibitem{GAP1} A.B. Aceves, Optical gap solitons: Past, present, and future; theory and experiments. Chaos \textbf{10}, 584--589 (2000)

\bibitem{KA} Y.S. Kivshar, G.P. Agrawal, \textit{Optical Solitons: From Fibers to Photonic Crystals} (Academic Press, San Diego, 2003)

\bibitem{OPT1} D. Mihalache, D. Mazilu, F. Lederer, B.A. Malomed, Y.V. Kartashov, L.C. Crasovan, L. Torner, Stable spatiotemporal solitons in Bessel optical lattices. Phys. Rev. Lett. \textbf{95}, 023902 (2005)

\bibitem{Chen} Z. Chen, M. Segev, D. N. Christodoulides, Optical spatial solitons: historical overview and recent advances. Rep. Prog. Phys. \textbf{75}, 086401 (2012)

\bibitem{REV4} L. Leblond, D. Mihalache, Models of few optical cycle solitons beyond the slowly varying envelope approximation. Phys. Rep. \textbf{523}, 61--126 (2013)

\bibitem{Tang} Y.F. Song, X.J. Shi, C.F. Wu, D.Y.Tang, H. Zhang, Recent progress of study on optical solitons in fiber lasers. Appl. Phys. Rev. \textbf{6}, 021313 (2019)

\bibitem{OPT3} X. Zhu, M.R. Beli\'{c}, D. Mihalache, D. Xiang, L. Zeng, Two-dimensional gap solitons supported by a parity-time-symmetric optical lattice with saturable nonlinearity and fractional-order diffraction. Opt. Laser Technol. \textbf{184}, 112426 (2025)

\bibitem{Cui} D. Mao, Z. Yuan, K. Dai, Y. Chen, H. Ma, Q. Ling, J. Zheng, Y.
Zhang, D. Chen, Y. Cui, Z. Sun, B.A. Malomed, Temporal and spatiotemporal soliton molecules in ultrafast fibre lasers. Nanophotonics \textbf{14}, 677--706 (2025)

\bibitem{Hulet} K.E. Strecker, G.B. Partridge, A.G. Truscott, R.G. Hulet, Bright matter wave solitons in Bose--Einstein condensates. New J. Phys. \textbf{5}, 73 (2003)

\bibitem{Brazh} V.A. Brazhnyi, V.V. Konotop, Theory of nonlinear matter waves in optical lattices. Mod. Phys. Lett. B \textbf{18}, 627--651 (2004)

\bibitem{BEC0} F.Kh. Abdullaev, Dynamics of bright matter wave solitons in a Bose Einstein condensate. Int. J. Mod. Phys. B \textbf{19}, 3415--3473 (2005)

\bibitem{BEC1} D.J. Frantzeskakis, Dark solitons in atomic Bose-Einstein condensates: from theory to experiments. J. Phys. A \textbf{43}, 213001 (2010)

\bibitem{Luca} L. Salasnich, Bright solitons in ultracold atoms. Opt. Quant. Electron. \textbf{49}, 409 (2017)

\bibitem{BEC2} B.A. Malomed, Creating solitons by means of spin-orbit coupling. EPL \textbf{122}, 36001 (2018)

\bibitem{BEC3} Y.V. Kartashov, D. A. Zezyulin, Stable multiring and rotating solitons in two-dimensional spin-orbit-coupled Bose-Einstein condensates with a radially periodic potential. Phys. Rev. Lett. \textbf{122}, 123201 (2019)

\bibitem{BEC4} Y.V. Kartashov, V.V. Konotop, Stable nonlinear modes sustained by gauge fields. Phys. Rev. Lett. \textbf{125}, 054101 (2020)

\bibitem{GAP2} Y.V. Kartashov, V.V. Konotop, F.K. Abdullaev, Gap solitons in a spin-orbit-coupled Bose-Einstein condensate. Phys. Rev. Lett. \textbf{111}, 060402 (2013)

\bibitem{GAP3} Y.V. Kartashov, F. Ye, V.V. Konotop, L. Torner, Multifrequency solitons in commensurate-incommensurate photonic moir\'{e} lattices. Phys. Rev. Lett. \textbf{127}, 163902 (2021)

\bibitem{GAP5} J. Chen, D. Mihalache, M.R. Beli\'{c}, J. Shi, D. Zhu, D. Deng, S. Qiu, R. Liao, X. Zhu, L. Zeng, Dark gap solitons in bichromatic optical superlattices under cubic-quintic nonlinearities. Chaos \textbf{34}, 113131 (2024)

\bibitem{GAP5a} L. Zeng, B.A. Malomed, D. Mihalache, and X. Zhu, One- and two-dimensional solitons under the action of the inverted cubic-quintic nonlinearity. Commun. Phys. \textbf{8}, 443 (2025)

\bibitem{GAP5b} X. Zhu, S. You, M.R. Beli\'{c}, D. Mihalache, J. Chen, L. Zeng, Bright soliton families in the nonlinear Schr\"{o}dinger equation with a bichromatic linear lattice. Opt. Express \textbf{33}, 46988--47000 (2025)

\bibitem{NL3} Y.V. Kartashov, B.A. Malomed, V.A. Vysloukh, L. Torner, Two-dimensional solitons in nonlinear lattices. Opt. Lett. \textbf{34}, 770--772 (2009)

\bibitem{AVK} H. Sakaguchi, B. A. Malomed, Solitons in combined linear and nonlinear lattice potentials. Phys. Rev. A \textbf{81}, 013624 (2010)

\bibitem{NL1} F.Kh. Abdullaev, Y.V. Kartashov, V.V. Konotop, D. A. Zezyulin, Solitons in $\mathcal{PT}$-symmetric nonlinear lattices. Phys. Rev. A \textbf{83}, 041805(R) (2011)

\bibitem{GAP4} L. Zeng, B.A. Malomed, D. Mihalache, J. Li, X. Zhu, Solitons in composite linear-nonlinear moir\'{e} lattices. Opt. Lett. \textbf{49}, 6944--6947 (2024)

\bibitem{NL4} Y.V. Kartashov, V.A. Vysloukh, L. Torner, Propagation of solitons in thermal media with periodic nonlinearity. Opt. Lett. \textbf{33}, 1774--1776 (2008)

\bibitem{SATU9} L. Zeng, J. Shi, M.R. Beli\'{c}, D. Mihalache, J. Chen, X. Lu, Y. Cai, J. Li, Multipole solitons in saturable nonlinear lattices. Nonlinear Dyn. \textbf{111}, 3665--3678 (2023)

\bibitem{SATU10} X. Zeng, M.R. Beli\'{c}, D. Mihalache, X. Lu, Y. Cai, J. Li, X. Zhu, L. Zeng, Two-dimensional soliton families in saturable quasi-nonlinear lattices. Opt. Express \textbf{33}, 33483--33493 (2025)

\bibitem{NLD1} X. Zhu, Y. Fan, M.R. Beli\'{c}, D. Mihalache, D. Xiang, L. Zeng, Optical dark solitons in purely cubic-quintic nonlinear lattices. Opt. Express \textbf{33}, 7205--7217 (2025)

\bibitem{NLD2} X. Zhu, M.R. Beli\'{c}, D. Mihalache, J. Li, L. Zeng, Dark solitons in purely quintic nonlinear lattices with defects. Chaos Solitons Fract. \textbf{200}, 116983 (2025)

\bibitem{NL6} J. Shi, L. Zeng, J. Chen, Two-dimensional localized modes in saturable quintic nonlinear lattices. Nonlinear Dyn. \textbf{111}, 13415--13424 (2023)

\bibitem{CS2} Y.V. Kartashov, V.V. Konotop, Solitons in Bose-Einstein condensates with helicoidal spin-orbit coupling. Phys. Rev. Lett. \textbf{118}, 190401 (2017)

\bibitem{CS3} L. Zeng, J. He, B.A. Malomed, J. Chen, X. Zhu, Spontaneous symmetry and antisymmetry breaking of two-component solitons in a combination of linear and nonlinear double-well potentials. Phys. Rev. E \textbf{110}, 064216 (2024)

\bibitem{CS3a} J. Chen, D. Mihalache, M.R. Beli\'{c}, W. Qin, D. Zhu, X. Zhu, L. Zeng, Dark gap soliton families in coupled nonlinear Schr\"{o}dinger equations with linear lattices. Nonlinear Dyn. \textbf{113}, 10307-10318 (2025)

\bibitem{VTS1} D. Mihalache, D. Mazilu, L. Torner, Stability of walking vector solitons. Phys. Rev. Lett. \textbf{81}, 4353--4356 (1998)

\bibitem{VTS2} J. Yang, Y. Tan, Fractal structure in the collision of vector solitons. Phys. Rev. Lett. \textbf{85}, 3624--3627 (2000)

\bibitem{VTS3} J. Yang, Interactions of vector solitons. Phys. Rev. E \textbf{64}, 026607 (2001)

\bibitem{VTS4} L. Salasnich, B. A. Malomed, Vector solitons in nearly one-dimensional Bose-Einstein condensates. Phys. Rev. A \textbf{74}, 053610 (2006)

\bibitem{VTS5} P. Balla, G.P. Agrawal, Nonlinear interaction of vector solitons inside birefringent optical fibers. Phys. Rev. A \textbf{98}, 023822 (2018)

\bibitem{VTS6} Y.V. Kartashov, Vector solitons in parity-time-symmetric lattices, Opt. Lett. \textbf{38}, 2600--2603 (2013)

\bibitem{VTS7} J.J. Garc\'{\i}a-Ripoll, V.M. P\'{e}rez-Garc\'{\i}a, E.A. Ostrovskaya, Y.S. Kivshar, Dipole-mode vector solitons. Phys. Rev. Lett. \textbf{85}, 82--85 (2000)

\bibitem{VTS8} W. Krolikowski, E.A. Ostrovskaya, C. Weilnau, M. Geisser, G. McCarthy, Y.S. Kivshar, C. Denz, B. Luther-Davies, Observation of dipole-mode vector solitons. Phys. Rev. Lett. \textbf{85}, 1424--1427 (2000)

\bibitem{VTS9} S. Stalin, M. Lakshmanan, Vector soliton molecules and their collisions. Phys. Rev. E \textbf{111}, 044205 (2025)

\bibitem{VTS10} A.S. Desyatnikov, Y.S. Kivshar, Necklace-ring vector solitons. Phys. Rev. Lett. \textbf{87}, 033901 (2001)

\bibitem{VTS11} J. Yang, D.E. Pelinovsky, Stable vortex and dipole vector solitons in a saturable nonlinear medium. Phys. Rev. E \textbf{67}, 016608 (2003)

\bibitem{VTS12} I.L. Garanovich, A.A. Sukhorukov, Y.S. Kivshar, M.
Molina, Surface multi-gap vector solitons. Opt. Express \textbf{14},
4780--4785 (2006)

\bibitem{VTS13} Y.V. Kartashov, F. Ye, L. Torner, Vector mixed-gap surface solitons. Opt. Express \textbf{14}, 4808--4814 (2006)

\bibitem{Panoiu} S. Lavdas, N.C. Panoiu, Theory of pulsed four-wave mixing in one-dimensional silicon photonic crystal slab waveguides. Phys. Rev. B \textbf{93}, 115435 (2016)

\bibitem{Bianc} M.F. Saleh, F. Biancalana, Soliton dynamics in gas-filled hollow-core photonic crystal fibers. J. Opt. \textbf{18}, 013002 (2016)

\bibitem{hetero} E. Istrate, E.H. Sargent, Photonic crystal heterostructures and interfaces. Rev. Mod. Phys. \textbf{78}, 455--481 (2006)

\bibitem{Notomi} M. Notomi, Manipulating light with strongly modulated photonic crystals. Rep. Prog. Phys. \textbf{73}, 096501 (2010)

\bibitem{Moiseyev}
N. Moiseyev, L.D. Carr, B.A. Malomed, Y.B. Band, Transition from resonances to bound states in nonlinear systems:
application to Bose-Einstein condensate. J. Phys. B: At. Mol. Opt. Phys. \textbf{37}, L193--L200 (2004).

\bibitem{MSOM} J. Yang, T.I. Lakoba, Universally-convergent squared-operator iteration methods for solitary waves in general nonlinear wave equations. Stud. Appl. Math. \textbf{118}, 153--197 (2007)

\bibitem{L0B} R.M. Caplan, R. Carretero-Gonz\'{a}lez, Numerical stability of explicit Runge-Kutta finite-difference schemes for the nonlinear Schr\"{o}dinger equation. Appl. Numer. Math. \textbf{71}, 24--40 (2013)

\bibitem{FCM} J.P. Boyd, \textit{Chebyshev and Fourier spectral methods} (Courier Corporation, 2001)

\end{thebibliography}
\end{document}